\documentclass[
reprint,
bibnotes,
 amsmath,amssymb,
 aps,
pra,
]{revtex4-2}

\usepackage{graphicx}
\usepackage{dcolumn}
\usepackage{bm}
\usepackage{siunitx}

\usepackage[utf8]{inputenc}
\renewcommand{\selectlanguage}[1]{}

\begin{document}

\preprint{APS/123-QED}

\title{Delay in electronic vortex states created by multiphoton ionization with single elliptically polarized laser pulses}

\author{Edward McManus$^1$}
\email{These authors contributed equally to this work}
\author{Phi-Hung Tran$^1$}
\email{These authors contributed equally to this work}
\author{Michael Davino$^1$}
\author{Tobias Saule$^1$}
\author{Van-Hung Hoang$^2$}
\author{Thomas Weinacht$^3$}
\author{ George Gibson$^1$}
\author{Anh-Thu Le$^1$}
\email{thu.le@uconn.edu}
\author{Carlos A. Trallero-Herrero$^1$}
\email{carlos.trallero@uconn.edu}
\affiliation{$^1$ Department of Physics, University of Connecticut, Storrs, CT 06268, USA\\
$^2$ J.~R. Macdonald Laboratory, Department of Physics, Kansas State University, Manhattan, Kansas 66506, USA\\
$^3$ Department of Physics and Astronomy, Stony Brook University, Stony Brook, New York 11794-3800, USA}


\date{\today}

\begin{abstract}
We show experimentally and theoretically that vortex-shaped structures in the photoelectron momentum distribution can be observed for atoms interacting with a single intense elliptically polarized laser pulse. Our analysis reveals that these spiral structures are the result of destructive interference of two dominant photoelectron vortex states, which are released into the continuum by strong-field ionization. An electron ``born" into one of those states is temporarily delayed near the atomic core by the combined atomic and laser potential, leading to fast changes in the phase delay with energy for photoelectrons in these vortex states. Our results open the door to studying electron dynamics of vortex states in strong-field ionization.

\end{abstract}

\maketitle


Optical vortices have been extensively studied over the past three decades since the pioneering work of Allen {\it et al.} \cite{allen_orbital_1992}, leading to numerous applications in optical tweezers, classical and quantum communication, and phase contrast imaging \cite{yao_orbital_2011,shen_structured_2020,forbes_structured_2021}.  Recently, beams of vortex electrons carrying well-defined orbital angular momentum (OAM) were proposed \cite{bliokh_optical_2017}. Subsequently, the generation and manipulation of such vortex electrons have attracted a great deal of attention in atomic physics,  due to potential applications in high-resolution electron microscopy and spectroscopy \cite{uchida_generation_2010,verbeeck_production_2010,bliokh_theory_2017,lloyd_electron_2017}. In fact, the OAM of the vortex electrons is expected to provide new information about the electronic and magnetic properties of the targets \cite{bliokh_theory_2017,lloyd_electron_2017}. 

In the context of strong-field and attosecond physics, the ultrafast formation of electronic vortex states in atoms was theoretically proposed a decade ago \cite{ngoko_djiokap_electron_2015} and experimentally observed in a few-photon transition in atomic potassium \cite{pengel_electron_2017}. However, so far the formation and observation of such vortex states required the use of multicolor and/or multi-ellipticity fields to induce the spiral pattern typical of vorticity. Therefore, one might have the impression that electronic vortex states are a rarity in strong-field physics, only attainable under quite stringent conditions for the driving electric field \cite{Yuan:pra2016}. %



Theoretically, the first prediction of electron vortex production by photoionization was made in 2015 by Djiokap {\it et al.} \cite{ngoko_djiokap_electron_2015}. They showed, based on time-dependent Schr{\"o}dinger equation (TDSE) simulations, that by using two time-delayed counter rotating circularly polarized (CCP) attosecond pulses to ionize helium, the photoelectrons are emitted with OAM determined by the helicity of the ionizing photon, and interfere, forming two armed spirals in the photoelectron momentum distribution (PED), with the time delay between the two pulses encoded in the angular rotation of the spiral arms. To our knowledge, the first experimental measurement of vorticity in photoelectrons was by Pengel {\it et al.} \cite{pengel_electron_2017,Pengel:pra2017} in the absorption of three photons in potassium with CCP pulse pairs. Their measurement revealed a six-way symmetric spiral PED at low energy, and a four-way symmetric distribution at high energy, corresponding to a population shift from the ground to an excited state with non-zero angular momentum. An extension of this technique to atomic sodium has also been reported \cite{Kerbstadt:NatComm2019,Kerbstadt:AdvPhys2019,Eickhoff:njp2020}. Following this experimental progress, theoretical analyses of vortex states in strong-field ionization have been reported \cite{Maxwell:Faraday2021,Kang:EPJD2021,Armstrong:pra2019,Tolstikhin:pra2019,Bazarov:pra2023,Planas:prl2022,Velez:pra18}.   

In this Letter, we investigate multiphoton ionization near the tunneling regime in Xe and H atoms by a \textit{single} elliptically polarized intense laser pulse. New measurements on Xe are compared with previous measurements on H \cite{trabert_nonadiabatic_2021} and new calculations for both atoms.  Experimentally, we measure the full 3D PED, which allows us to observe the spiral pattern in the plane of polarization. Furthermore, we show that the experimental results can be reproduced by two separate theoretical calculations, demonstrating that the spiral patterns are due to interference between states of different OAM which are emitted into the continuum with different delays. We thus argue that electronic vortex states are more ubiquitous than originally thought, but that observation is difficult because of the need of high-resolution 3D PEDs. Additionally, the formation of spiral patterns can be used as a temporal reference to clock time evolution at sub-cycle time scales without the need for introducing carrier envelope phase stability or attosecond pulses. Our measurements are similar in nature to the so-called ``atto-clock" or delayed ionization \cite{Eckle:NatPhys2008,Eckle:science2008,Schultze:science2010,Kheifets:prl2010,Klunder:prl2011,Dahlstrom:jpb2012,Landsman:Optica14,Landsman:PhysRep2015} but between two well-defined vortex states.

\begin{figure*}
    \centering
    \includegraphics[width=0.95\linewidth]{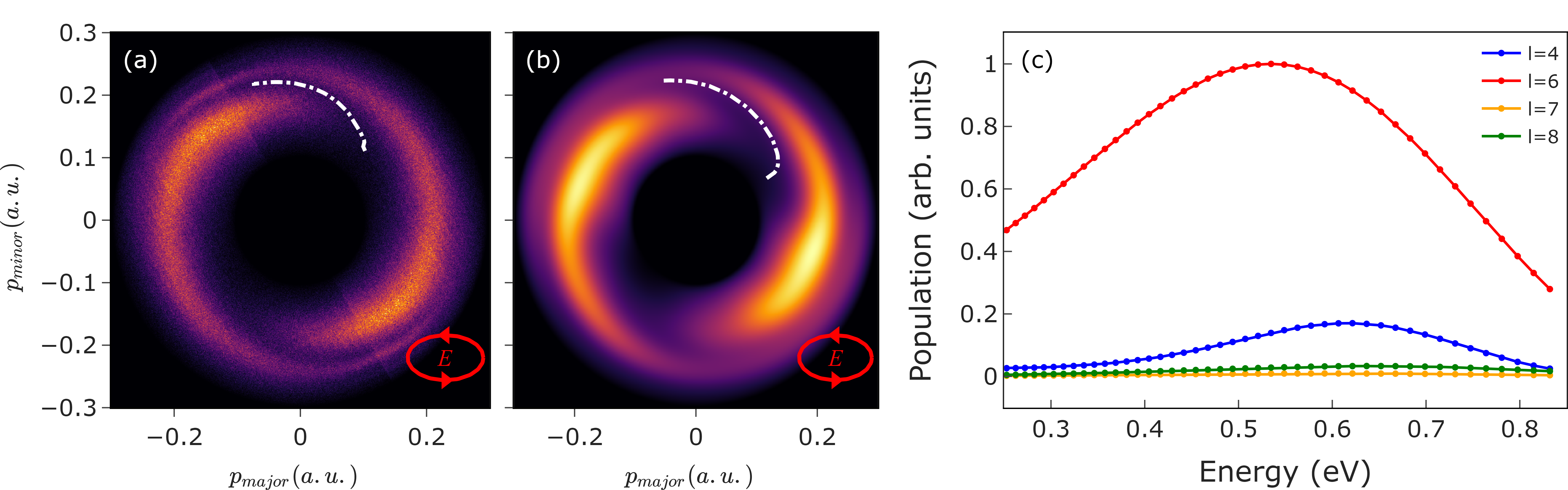}
    \caption{(a), (b) The 2D polarization-plane PED from Xe by an elliptically polarized laser pulse with $\varepsilon=0.6$. Experimental data (a) shows the raw measurement of the first ATI ring integrated over a thin slice (width 0.1 a.u.) centered on the polarization plane. TDSE calculations (b) are shown along the polarization plane for the same conditions as the experimental data. The white line in each plot shows the location of the retrieved minima. The theory is used for calibration of the experimental momentum. In (c), the decomposition of the TDSE results in the vortex states defined in Eq.~(\ref{eq:vortex}) is shown for the ionization from the dominant $m=-1$ orbital. Notice that in (c) populations are shown vs energy with $E=p^2/2$.}
    \label{fig:peds}
\end{figure*}
Our experimental setup used has been described elsewhere \cite{davino_plano-convex_2023}. In short, Ti-Sapphire laser pulses (80 fs, 47 $\mu$J), are focused into an effusive molecular beam in a 3D velocity map imaging (3DVMI) apparatus, with an intensity at the focus of $\approx2\times10^{13}$  W/cm$^2$ (Keldysh parameter $\gamma\approx 2$), determined by the pulse energy, duration, and focal spot and confirmed by comparison with TDSE simulations. The 3DVMI is capable of measuring the time of flight (ToF) of electrons on the propagation axis by using a timepix3 camera \cite{cheng_3d_2022,davino_plano-convex_2023}. All 2D PED results presented here are for photoelectrons measured in the plane of polarization defined by the electron ToF axis and the axis perpendicular to both the propagation and ToF axes. In addition, the apparatus can measure the ion time of flight (iToF) shot-by-shot in coincidence with the electrons which we use to gate the electrons. Such coincidence (Xe$^+$ + e$^-$) data analysis provide very clean above-threshold ionization (ATI) spectra at modest vacuum (10$^{-8}$ torr). Measurements were taken for a variety of ellipticities ($\varepsilon$), defined as the experimental ratio of the electric field amplitude between the major and minor axis, with the data presented in this letter taken at $\varepsilon=0.6$, although the interpretation presented is consistent across a wide range of ellipticities, down to $\varepsilon=0.2$. A symmetry analysis of similar data was presented in our earlier work \cite{Davino:prapp2025}\\


The 2D PED measured in the polarization plane of Xe by intense left-handed elliptically polarized lasers with $\epsilon=0.6$ is shown in Fig.~\ref{fig:peds}(a). Here, we only focus on the first ATI ring (see the Supplemental Material for the full experimental PED). The most striking feature of this PED is the appearance of a spiral interlocking structure. In fact, this structure resembles the vortex-shaped PED observed in multiphoton ionization of potassium by Pengel {\it et al.} \cite{pengel_electron_2017}, and also theoretically predicted  by Djiokap {\it et al.} \cite{ngoko_djiokap_electron_2015} for single-photon ionization of helium. We emphasize that in our experiment only a single elliptically polarized pulse was used, in contrast to those earlier works. The corresponding TDSE result, shown in Fig.~\ref{fig:peds}(b), reproduces very well the interlocking structure of the measured PED. The theoretical PED includes contributions from all $m$ in the 5p shell, i.e. $m=0$ and $\pm 1$ of Xe(5p), and the averaging of the intensity variation in the laser focus has been carried out for the peak intensity of $2.2\times 10^{13}$ W/cm$^2$. In the following, we will show that the minimum positions of the first ATI ring, indicated as the dashed-dotted line in Fig. 1(a) and 1(b), contain dynamic information about the ionization process. 

To provide insight into the PED structures, we expand the TDSE solution at the end of the laser pulse, in basis set of vortex states. More specifically, this can be written in momentum space as \cite{lloyd_electron_2017,bliokh_theory_2017,Kang:EPJD2021,Maxwell:Faraday2021}
\begin{eqnarray}
    \psi(p, \phi)= \sum_\ell
    C_\ell(p) e^{i \ell \phi},\nonumber \\
    C_\ell(p) = |C_\ell(p)|e^{i \varphi_\ell(p)}   
    \label{eq:vortex}
\end{eqnarray}
with $\phi$ being the azimuthal angle, $\ell$ -- the azimuthal quantum number of the orbital angular momentum (OAM) (also known as the topological charge, or winding number), and $C_\ell$ -- the complex expansion coefficients in the vortex basis set with its phase $\varphi_\ell$. 

The probability distribution $|C_\ell(E)|^2$ is shown in Fig.~\ref{fig:peds}(c) for the four dominant terms in the expansion starting from the Xe(5p$_{m=-1}$), at a single laser intensity of $2.2\times 10^{13}$ W/cm$^2$. Clearly, the pathways leading to $\ell=6$ and $\ell=4$ dominate the first ATI ring. This can be understood as follows. An elliptically polarized pulse can be represented as a superposition of two pulses -- left- and right-handed circularly polarized -- with different weights, depending on the pulse helicity. The first ring can be reached by an electron from the $5p$ shell by absorbing 9 photons, each photon contributing a $+1$ (or $-1$) unit of angular momentum along its propagation direction, for left-handed (right-handed) polarization. Photoelectrons launched into the first ATI continuum will have energy $E = 9\hbar\omega - (I_p + U_p) \approx 0.6$ eV, or momentum $p \approx 0.2$ a.u., consistent with Fig.~\ref{fig:peds}. Therefore, only vortex states with even $\ell\le 8$ are populated [see Fig.~1(c) for an example of a negligible population of the vortex state with an odd $\ell=7$]. The small population of the vortex state with $\ell=8$ is attributed to rather small ellipticity of the laser with $\varepsilon=0.6$ \cite{Bashkansky:prl1988} -- also see below an example of hydrogen with a larger $\varepsilon=0.85$. Thus, the interlocking structure shown here can simply be explained as destructive interference between two most dominant vortex wave packets with $\ell=6$ and $\ell=4$. Note that contributions from electrons in $m=+1$ and $m=0$ orbitals are significantly weaker \cite{Barth:pra2011,Herath:prl2012,Barth:pra2013}, resulting in nearly the same minimum positions in the first ATI ring of the total yield as compared to that of the partial yield from $m=-1$. The location where the destructive interference is most notable is shown as white lines in Fig.~\ref{fig:peds} (a) and (b). We will show below that this interlocking spiral structure gives us access to the phase difference between those two vortex states. 


\begin{figure}[tb!]
    \centering
    \includegraphics[width=0.85\linewidth]{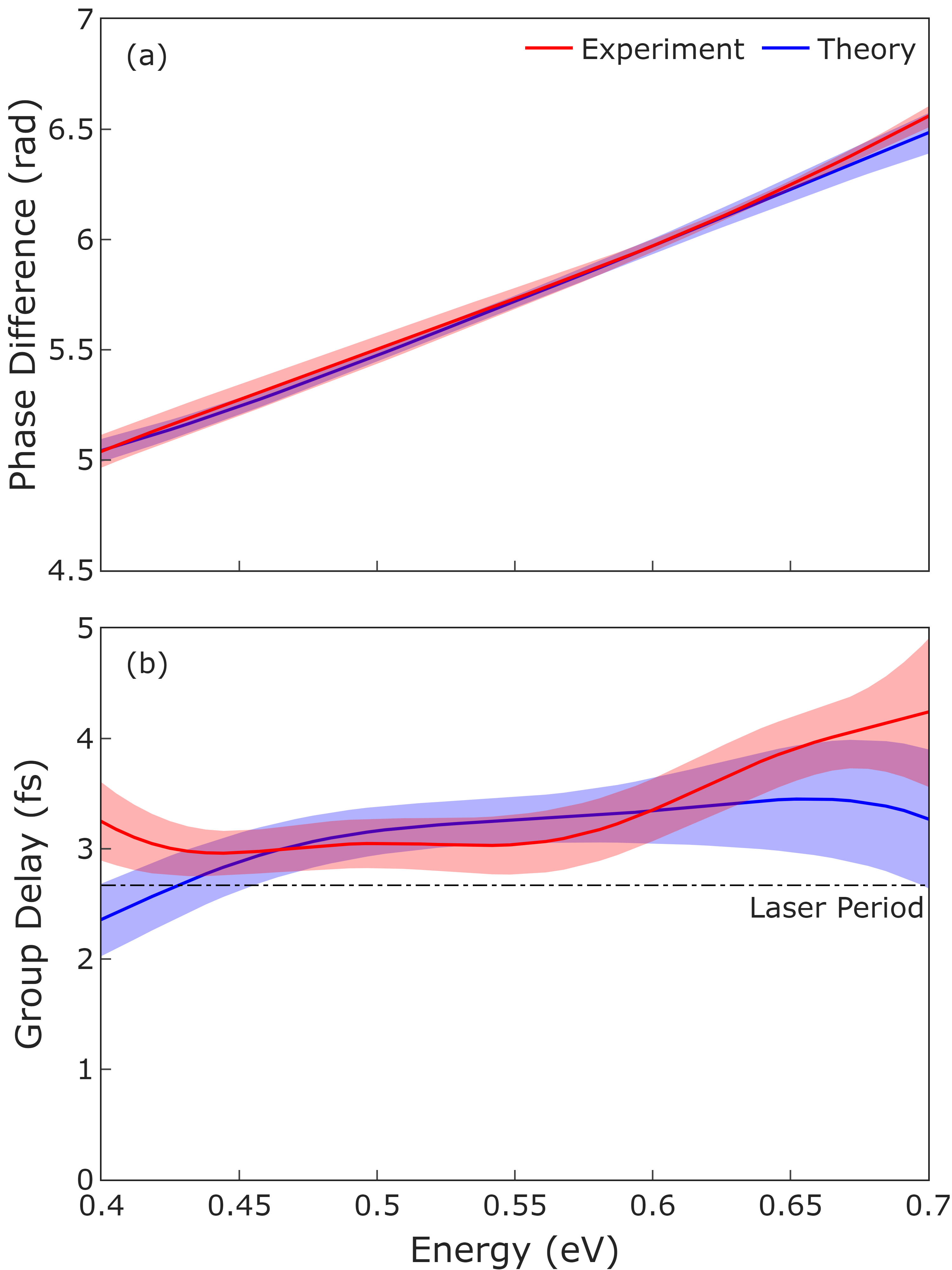}
    \caption{(a) Comparison of experimentally retrieved (in red) phase difference $\Delta\varphi_{46}$ with TDSE result (in blue). (b) The same as in (a), but for group delay $\tau_g$. The laser period of 2.7 fs (black) is noted as a dash-dotted line in panel (b). Shaded regions indicate the uncertainty, discussed further in section II of the Supplementary Material.}
    \label{fig:delays}
\end{figure}

For each momentum $p$ in the 2D PED, the azimuthal position $\phi_{min}$ of the destructive interference between the photoelectron wave packets with $\ell=4$ and $\ell=6$ is related to the phase difference of these vortex states $\Delta\varphi_{46}=\varphi_{\ell=4} - \varphi_{\ell=6}$ by an equation $\phi_{min}=\Delta\varphi_{46}/2 + \pi/2$ (see the Supplemental Material). For our purpose, we use this mapping equation to retrieve the phase difference $\Delta\varphi_{46}$ from the minimum position of the measured PED and the TDSE and the result is shown in Fig.~\ref{fig:delays}(a). The agreement between experiment and theory is excellent, with $\Delta\varphi_{46}$ varying between 5 and 6.5 rad across the ATI peak. Additionally, a comparison of the retrieved group delay, defined as $\tau_g=\hbar\frac{d\Delta\varphi_{46}} {dE}$, is also shown in Fig. \ref{fig:delays}(b) for theory and experiment. Both results again are in good agreement and show $\tau_g$ close to the laser period, indicating the role of the laser electric field in the ionization dynamics. These interpretations have been confirmed by our semiclassical simulations for the hydrogen target, as shown below.

\begin{figure}[tb!]
\begin{center}
\includegraphics[width=0.9\linewidth]{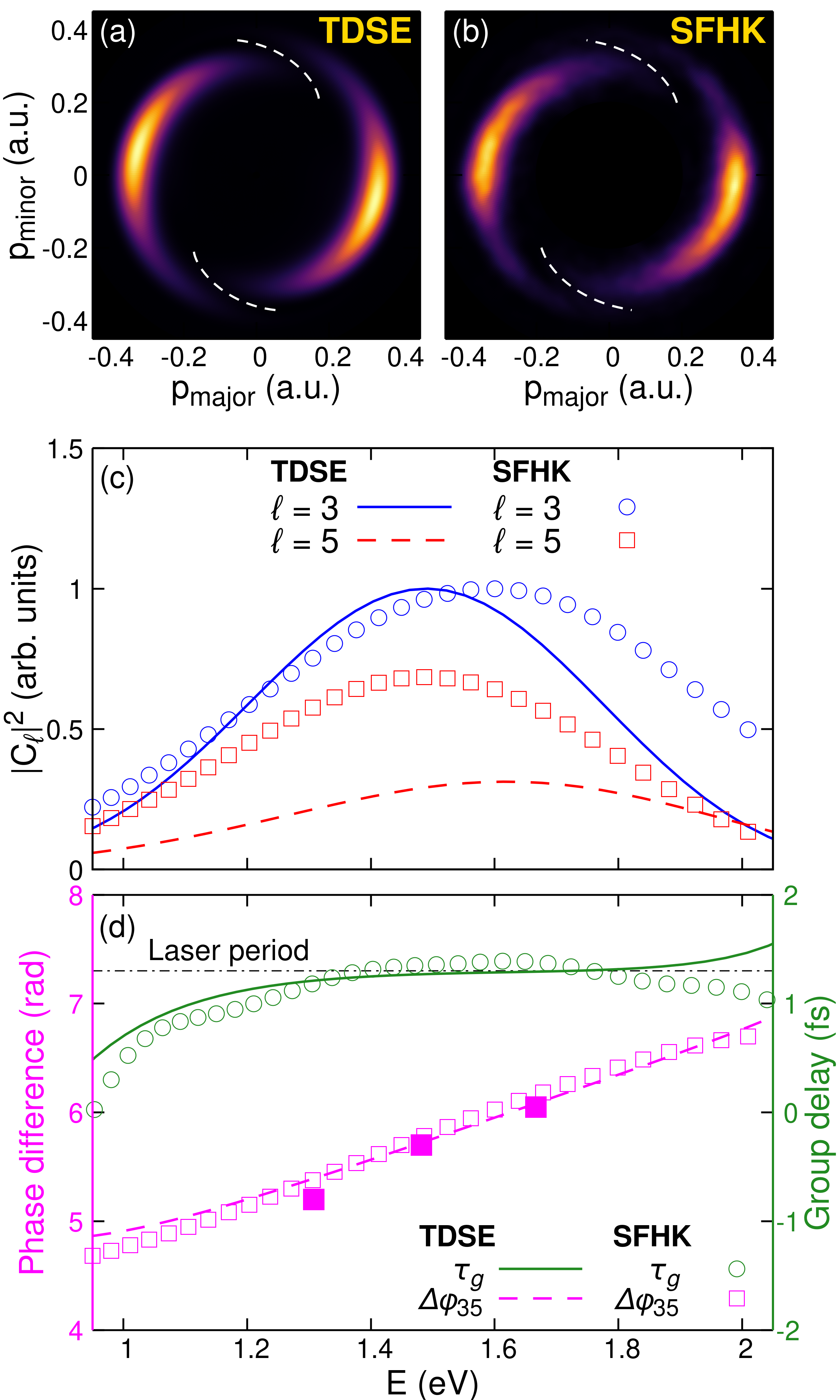}
\caption{The theoretical PED in the polarization plane from atomic hydrogen by 390 nm elliptically polarized light, calculated with the TDSE (a) and the SFHK (b). The white dash lines show the position of the minimum. (c) Decomposition of the population of the electrons in vortex states with OAM number $\ell=3,5$. (d) Phase difference $\Delta \varphi_{35} = \varphi_{\ell=3}- \varphi_{\ell=5}$ (magenta) between the two dominant $\ell$, for TDSE (dotted line) and SFHK (solid line) simulations, along with the corresponding group delay (green). The laser period (black line) at 1.33 fs is noted. Solid magenta squares are the phase difference using classical trajectories for $E=1.31$, $1.48$, and $1.67$~eV, see Fig.~\ref{fig:trajectories}.}
\label{fig:Hydrogen_PED}
\end{center}
\end{figure}

To corroborate the nature of the vortex state interference, we now employ our recently developed strong-field Herman-Kluk propagator method (SFHK) \cite{tran_quantum_2024}. This semiclassical method has been shown to be capable of providing accurate PED \cite{tran_quantum_2024, Hasan:arxiv2025}. To simplify the analysis, we will use atomic hydrogen as a target instead of Xe. Here, only one orbital (i.e., the $1s$ orbital) contributes to the signal in the PED. This choice is also motivated by the recent experiment on strong-field ionization of atomic hydrogen by Trabert {\it et al.} \cite{trabert_nonadiabatic_2021} using elliptically polarized laser pulses at the 390-nm wavelength, with ellipticity of $\varepsilon=0.85$. In fact, both their measured PED and TDSE results showed some signature of a vortex-shaped structure. However, the main focus of that article was on the origin of the ring-to-ring offset angles. Furthermore, the fast dependence of the offset angle within single ATI peaks (which is the main feature of the vortex-shaped structure), was speculated as due to focal averaging. In contrast, our calculated 2D PEDs from the TDSE and SFHK, using the same laser parameters as in the Trabert {\it et al.} experiment, but at a {\em single} laser intensity of $5\times 10^{13}$ W/cm$^2$, both show a clear vortex-shaped structure -- see Fig.~\ref{fig:Hydrogen_PED}(a) and (b). We found that focal averaging further enhances this feature. We emphasize the excellent agreement between the two theoretical methods, including the minimum position of the yields (the white dotted lines).

Following the same analysis as in the case of Xe, we project the 2D PED for the first ATI ring in the vortex state basis Eq.~(\ref{eq:vortex}).The populations $|C_\ell(E)|^2$ are shown in Fig.~3(c) for both TDSE and the SFHK approaches. In both methods, the populations are dominated by $\ell=3$ and $\ell=5$ states, consistent with the fact that an absorption of at least five photons is needed for the H(1s) electron to reach the continuum. Indeed, since a right-handed elliptically polarized pulse (with $\varepsilon=0.85$) can be represented as a linear superposition of a left-handed circularly polarized pulse (with a smaller weight) and a right-handed circularly polarized pulse (with a larger weight), the most likely pathway would be absorbing four right-handed photons and one left-handed photon. In this pathway, the final photoelectron would have the angular momentum of $3\hbar$ along the $z$-axis (i.e., the topological charge $\ell=3$). Another likely pathway is for the 1s electron to absorb all five right-handed photons to end up in the continuum with $\ell=5$. There is an overall good agreement between the TDSE and SFHK methods, although the SFHK overestimates the population of $\ell=5$ and there are some small shifts in the peak positions of the SFHK populations as compared to that of the TDSE.


As in the case of Xe, the minimum structure in the first ATI ring is due to a destructive interference between the most dominant contributions from the $\ell=3$ and $\ell=5$ states. It is also possible to extract the phase difference $\Delta\varphi_{35}=\varphi_{\ell=3} - \varphi_{\ell=5}$ between the two contributing states and group delay $\tau_g$ between $\ell=3$ and $\ell=5$ vortext states, see Fig.~\ref{fig:Hydrogen_PED}(d). We note a remarkably good agreement between the TDSE and the SFHK results in the phase difference, indicating the validity of the SFHK method. Group delay values do diverge at the edges of the ATI ring; more details will be presented elsewhere.

\begin{figure}[tb!]
\begin{center}
\includegraphics[width=0.8\linewidth]{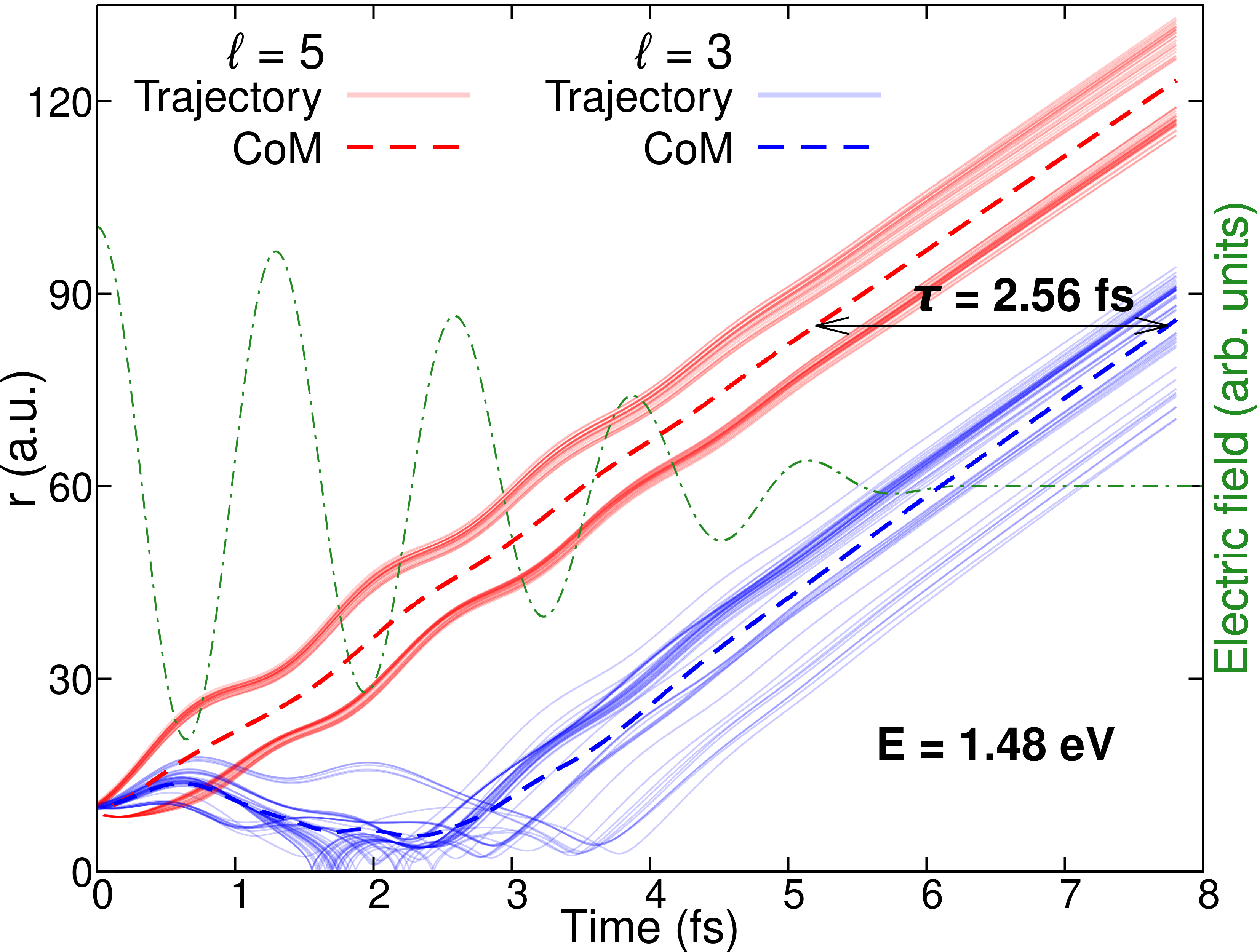}
\caption{
Sample classical trajectories for photoelectrons with $E=1.48$~eV, with OAM number $\ell=3,5$. The center of mass (CoM) of the trajectories is also shown. The majority of the paths with $\ell=3$ got trapped temporarily near the atomic core by the combined atomic and laser electric field potentials for almost two laser periods 2T$_{laser}=2.66$ fs, leading to a separation between the CoM of the $\ell=3$ and $\ell=5$ pathways of $\tau=2.56$ fs. The phase difference calculated with the ``retrieved" $\tau$ is shown in Fig.~\ref{fig:Hydrogen_PED}(d), together with other two energies (1.31 and 1.67~eV) as solid squares.}
\label{fig:trajectories}
\end{center}
\end{figure}

Thanks to the semi-classical nature of the SFHK we are now in a position to provide a further intuitive picture on the formation of the vortex destructive interference pattern. We show in Fig.~\ref{fig:trajectories} sample classical trajectories for electrons emitted with final kinetic energy $E=1.48$ eV (which is near the center of the first ATI peak), with {\it final} $\ell = 3$ and $\ell=5$. These trajectories are first simulated using the strong-field approximation (SFA) in the ionization step. The electron is then propagated in the combined field of the Coulomb potential and the laser electric field. Note that within the SFHK, typically millions of these trajectories are used to generate the continuum wavefunction with the Herman-Kluk propagator \cite{tran_quantum_2024}. Remarkably, the $\ell=5$ electrons generally escape the atom very quickly after ``birth" in the continuum, whereas the $\ell=3$ electrons seem to be temporarily delayed near the atomic core by the effective potential and roam for some time before eventually escaping. To quantify this effect, we calculate the center of mass (CoM) of each type of trajectories and the results are shown as thick dashed lines. After the end of the laser, these two lines are separated by a constant value, with the CoM of the $\ell=3$ trajectories delayed by about 2.56 fs compared to that of $\ell=5$, almost two times the laser period of $T=1.33$ fs. 
This phase delay can be converted into the phase difference by $\tau_p=\hbar\frac{\Delta\varphi_{35}} {E}$ and is shown in Fig.~\ref{fig:Hydrogen_PED}(d) as solid squared, for $E=1.48$ eV together with $E=1.31$ and 1.67 eV (see the SM). Clearly, all three methods agree well. Thus, this classical picture shows that two dominant classes of trajectories with corresponding OAM numbers $\ell=3,5$ interfere when electron is ``launched" in the continuum two laser periods apart.

The discussion above paints a clear answer on the conditions for the formation of a vortex in strong field ionization: due to the difference in angular momentum, electrons with lower $\ell$ are launched into the continuum significantly later relative to electrons with higher $\ell$, and in addition, there must be a non-zero group delay (aka group velocity dispersion) across the energies of the ATI ring. So far we have only investigated the vortex patterns in systems with only two dominant states in expansion Eq.~\ref{eq:vortex}, but it should not be limited to only two, although it can be harder to observe and analyze. 

To conclude, we measured the polarization-plane strong-field PED in a 3D VMI showing the formation of spiral patterns indicative of an interference of vortex states. Experimental observation of vortex states interference was achieved with a single elliptically polarized pulse at wavelength of 800 nm in Xe or 390 nm in H. This is in contrast with previous results where either bicolor or counter-rotating pulses are needed to observe vortex states. We show that experimental results can be accurately reproduced by our theoretical calculations through two models: numerical solution of the TDSE and a trajectory-based semiclassical SFHK method. In both cases, the results are expanded in a vortex-states basis set, demonstrating that only two states are the main contributors for both atoms. The spiral interlocking pattern is formed as a result of interference between those two vortex states when there is significant group velocity dispersion and the trajectories from the two contributing winding numbers $\ell$ are significantly delayed from one another. The results could open the door for holographic measurements of photoelectrons since there is always one direct electron that serves as a ``reference" in respect to which a second electron interferes after accumulating information on the target core potential. More broadly, there has been significant interest lately on the impact of vortex states in superconductivity \cite{yeh_structured_2025} and quantum matter in general \cite{quinteiro_rosen_interplay_2022} when driven by non-gaussian light modes, as well as ultrafast imaging of electronic coherences in molecules \cite{Wu:prl2025}. This deeper understanding could lead to new quantum control strategies \cite{croitoru_toward_2022}.

Theoretical simulations by P.H.T. and A.T.L. were supported by DOE BES, Chemical Sciences, Geosciences, and Biosciences Division Grant No. DE-SC0023192. Experiments were performed under DOE BES, Chemical Sciences, Geosciences, and Biosciences Division Grant No. DE-SC0024508.

\bibliographystyle{apsrev4-2}
\bibliography{apssamp,carlos_zotero,Phi_Bib}

\end{document}